# Lithium Niobate Michelson Interferometer Modulator on Silicon-On-Insulator Platform

Mengyue Xu,[1] Wenjun Chen,[1] Mingbo He,[1] Xueqin Wen,[2] Ziliang Ruan,[2] Lifeng Chen[1,*], Liu Liu[2,*], Siyuan Yu,[1] Xinlun Cai[1,*]

[1] *State Key Laboratory of Optoelectronic Materials and Technologies and School of Electronics and Information Technology, Sun Yat-sen University, Guangzhou 510000, China*
[2] *South China Academy of Advanced Optoelectronics, South China Normal University, Guangzhou 510000, China*
*[*] chenlf37@mail.sysu.edu.cn , liu.liu@coer-scnu.org, caixlun5@mail.sysu.edu.cn*

**We propose and demonstrate a hybrid silicon and lithium niobate Michelson Interferometer Modulator (MIM) with enhanced modulation efficiency compared to a Mach-Zehnder modulator. The modulator is based on seamless integration of a high-contrast waveguide based on lithium niobate—a popular modulator material—with compact, low-loss silicon circuitry. The present device demonstrates a modulation efficiency as high as 1.2 V·cm and a low insertion loss of 3.3 dB. The 3dB electro-optic bandwidth is approximately 17.5 GHz. The optical eye diagrams, operating at 32 Gbit/s and 40 Gbit/s, with measured dynamic extinction ratios at 8 dB and 6.6 dB respectively. The present device avoids absorption loss and nonlinearity in conventional silicon modulators and demonstrates highest modulation efficiency in LN modulators, showing potential in future optical interconnects.**

http://dx.doi.org/10.1364/OL.99.099999

Silicon photonics on the silicon-on-insulator (SOI) platform has emerged as the leading technology for optical interconnect due to the possibility of low-cost and high-volume production of photonic integrated circuits (PICs) in CMOS foundries [1-3]. However, optical modulations in the silicon material mainly rely on free-carrier plasma dispersion effect, which leads to inevitable absorption losses, nonlinear voltage response and temperature sensibility [4, 5]. Lithium niobate ($LiNbO_3$, LN) shows potentials for realizing high performance electro-optic (EO) modulators due to its physical properties: large EO coefficient (30 pm/V), strong Pockles effect, wide bandgap (wide transparency window) and good temperature stability [6]. Nevertheless, commercial bulk LN modulators based on indiffused or proton-exchange waveguide are suffering with low refractive index contrast, low EO modulation efficiency (typically voltage-length product ($V_\pi \cdot L$)> 10 V·cm) [7], and difficult to integrate.

Lithium niobate on insulator (LNOI) has been reported as a promising platform for photonic integrated devices [7-11]. A typical cross-section of a LNOI photonic waveguide is less than 1μm², which leads to a small mode size and tight mode confinement. As a result, LNOI-based modulator allows for a good overlap between optical and electrical fields, and enhanced EO modulation efficiency. Recently, the integrated LN modulator has shown low loss, low drive voltage and large bandwidth [12]. An alternative approach, i.e., heterogeneous integration of LN membranes onto SOI photonic integrated circuits, has also attracted considerable interests [13-18]. The silicon/LN material system combines the scalability of silicon photonics with excellent modulation performance of LN. More recently, by etching the LN membrane, we have demonstrated a Mach-Zehnder modulator (MZM) based on the heterogeneous silicon/LN platform with low loss and large modulation bandwidth [19].

In this letter, we propose and demonstrate a Michelson interferometer modulator (MIM) based on the heterogeneous silicon/LN platform (Fig. 1(a)). In contrast with the travelling wave MZM structures, light wave is reflected by the reflective mirrors at the end of both arms, and the interaction length between the light wave and modulating electrical field doubles [20-23]. Figure 1(a) shows our heterogeneous silicon/LN MIM design. The MIM consists of a bottom silicon waveguide layer, a top LN waveguide layer and vertical adiabatic couplers (VACs) which transfer the optical power between the two layers. The top LN waveguides, formed by dry-etching of an X-cut LN thin film, serve as phase modulators where EO interactions occur. The bottom SOI circuit supports all other passive functions, consisting of a 3dB multimode interference (MMI) coupler that split and combine the optical power, two loop mirrors that serve as broadband reflectors, and two grating couplers for off-chip coupling. The VACs, which were formed by silicon inverse tapers and superimposed LN waveguides, serve as interfaces to couple light up and down between the two layers. A mode calculation result (using finite difference eigenmode (FDE) solver, Lumerical Mode Solution [24]) indicates the optical mode transferring 99.9% energy from the modulation region (mode 'A') to the beginning of VAC (mode 'B'), and vice versa. The continuous wave (CW) laser couples into the waveguide from one of the grating couplers. The RF signal is applied to ground-signal-ground (GSG) electrodes, and modulated light is detected at another grating coupler.

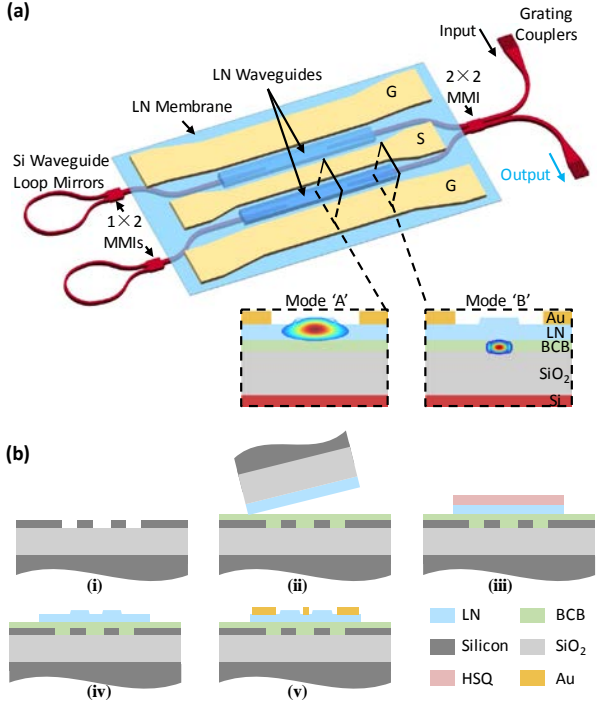

Fig. 1. (a) 3D chart of a heterogeneous silicon/LN MIM, insets are cross-section views of calculated optical TE0 mode in (mode 'A') the modulation region and (mode 'B') the beginning of VAC. (b) Device fabrication process.

The device fabrication process is shown in Fig. 1(b). Firstly, silicon grating couplers, MMIs, waveguide loop mirrors and inverse tapers are patterned by SOI processing including e-beam lithography (EBL) and dry etching. Then a commercially available X-cut LN membranes on insulator (LNOI) wafer from NanoLN is then flipped and bonded onto the patterned silicon wafer covered by benzocyclobuten (BCB). Removing the silicon substrate and oxide layer of the LNOI wafer leaves a stack of Si/BCB/LN film on the host substrate. Afterwards, the top LN waveguide, serving as phase modulators where Pockels effect occur, was patterned by e-beam lithography and dry etching process. Finally, a pair of travelling-wave electrodes, configured in a ground–signal–ground form, are fabricated directly on the LN layer by liftoff process.

Figure 2 shows the calculated $V_\pi \cdot L$ with optimization design of a LN etching depth and the corresponding gap between LN waveguide and electrode. The goal of the design is to enable high modulation efficiency with low metal absorption loss (here we target the total loss to 0.3 dB/cm). The electric field distribution ($E_z$) is simulated in a commercial software (COMSOL Multiphysics [25]). The EO refractive index change ($\Delta n$) in the LN material results in the change of the effective index of the TE$_0$ guided mode ($\Delta n_{ef}$) in the etched LN waveguide. Assume optical wave propagates along y-axis and the direction of the applied electrical filed is along z-axis on an X-cut LN wafer, the $\Delta n$ in the LN waveguide with a fixed applied voltage of 1 V is calculated by

$$\Delta n = E_z n_e^3 r_{33}, \qquad (1)$$

where $n_e$ is the LN refractive index and $r_{33}$ is the highest EO coefficient of LN. Then we use an optical FDE solver (Lumerical Mode Solution[24]) to solve the $\Delta n_{eff}$ for the fundamental TE mode. The single arm phase change of MIM in push-pull configuration is calculated by

$$\Delta \varphi = \frac{V}{V_\pi}\frac{\pi}{2} = \beta L = \frac{2\pi}{\lambda_0}\Delta n_{eff} L, \qquad (2)$$

where V is the fixed applied voltage of 1V, L is the EO interaction length in the MIM (twice the length of single modulation arm), $\beta$ is the propagation constant, $\lambda_0$ is the operating wavelength of 1550 nm. Therefore, we can calculate the value of $V_\pi \cdot L$ through solving $\Delta n_{ef}$ for $TE_0$ mode at a wavelength of 1550 nm. Figure 2 shows the lowest calculated $V_\pi \cdot L$ value is 1.05 V·cm when the top width of LN waveguide is 1 μm, and the correspondent etching depth of LN membrane is 0.3 μm. The gap between the edge of LN waveguide and the electrode is 1.8 μm. In practice, we fabricated LN ridge waveguides with etching depth of 0.18 μm, and the gap between the LN waveguide and the electrode is 2.75 μm. Our design has a modulation efficiency $V_\pi \cdot L$ of 1.2 V·cm which is not far away from the optimal situation.

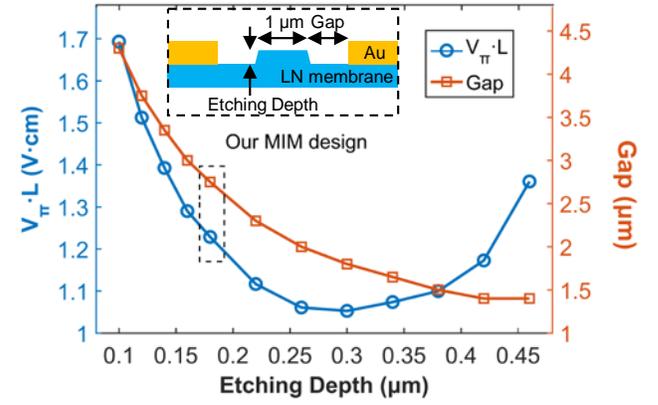

Fig. 2. The calculated $V_\pi \cdot L$ for MIM as a function of the LN etching depth (blue circle line). The corresponding smallest gap between the LN waveguide and the electrode is also depicted (red square line). The inset is the cross-sectional view of the modulation region to define the gap and the LN waveguide etching depth.

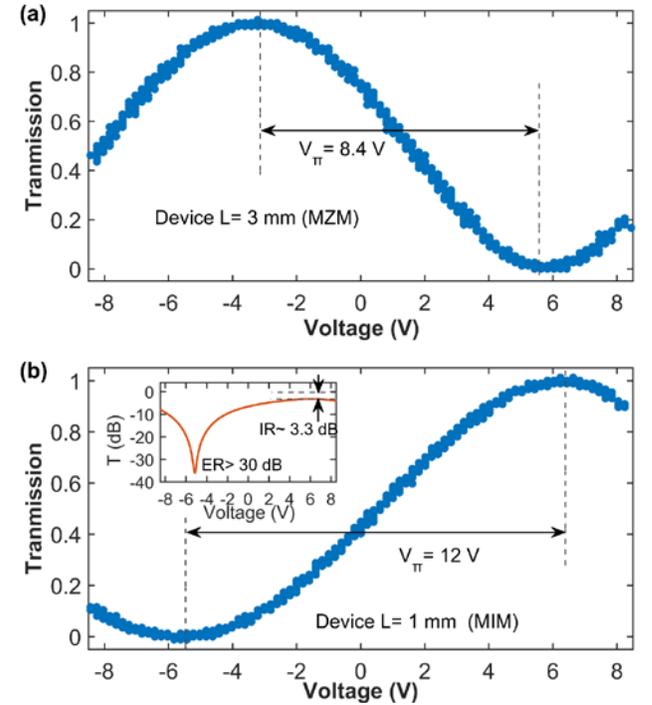

Fig. 3. Normalized optical transmission of (a) a 3-mm-long MZM and (b) a 1-mm-long MIM as a function of the applied voltage. The inset is the transmission normalized to grating couplers and shown in log scale.

To compare the performance of modulation efficiency and modulation bandwidth between MZM and MIM, we adopt the above-mentioned design and fabrication process for a 1-mm-long MIM and 3-mm-long MZM. Firstly, we performed half wave voltage ($V_\pi$) measurements with a 100-kHz triangular voltage sweep [18] for both modulators to evaluate modulation efficiency. Note that all measurement using 1550 nm wavelength as source center wavelength. The measured $V_\pi$ of a 3-mm-long MZM is 8.4 V (shown in Fig. 3.(a)), corresponding to $V_\pi \cdot L$ of 2.52 V·cm. Figure 3(b) shows the $V_\pi$ measured from the 1mm-long MIM device is 12 V, corresponding to $V_\pi \cdot L$ of 1.2 V·cm that lower than half of $V_\pi \cdot L$ in the 3mm-long MZM. This indicates the MIM has doubled modulation efficiency compare to the MZM as expected. The inset of Fig. 3(b) is the log-scaled transmission as a function of applied voltage, which shows the DC extinction ratio (ER) of >30 dB and the insertion loss of ~ 3.3 dB for 1mm MIM device. The total fiber-to-fiber loss is measured as 12.4 dB, including the coupling loss for per grating coupler (~4.7dB) and 0.13 dB loss for per VAC. We believe this $V_\pi \cdot L$ value of our MIM is the lowest amongst all previously reported thin-film LN devices [12, 13, 15, 16].

We then use a vector network analyzer (VNA, Agilent N5227A) to characterize the small signal EO bandwidth (S21 parameter) of our MIM and MZM device. Losses of RF cables and microwave probes (GGB 67A) are subtracted by using short-open-load-thru (SOLT) calibration standard with calibration substrates (GGB CS-5). The optical modulated signal is pre-amplified by an erbium-doped fiber amplifier (EDFA, Amonics AEDFA-PKT-DWDM-15-8-FA) and detected by a 70 GHz EO bandwidth photodiode (PD, FINISAR XPDV3120R). The electrode termination is loaded with 50 Ω. The EO bandwidth of our modulators are obtained by deducting the know EO $S_{21}$ of the PD. Figure 4 demonstrates the measurement result of electro-electro (EE) and electro-optic ($S_{21}$ parameter) response as a function of the input frequency sweep (blue solid line) in a 1-mm-long MIM and a 3-mm-long MZM. The measured EE $S_{21}$ is well above -6.4 dB point until 67 GHz. This indicates low losses of RF signal in transmission line for both modulators. The measured -3dB EO bandwidth of MIM is about 17.5 GHz (see in Fig. 4(b)) when the modulator biased at quadrature point (~0.5 V). For a MZM with same RF electrode design, it shows EO bandwidth above 67 GHz when biased at 1.6 V. Because of same RF electrode design for MIM and MZM, both modulators have same impedance and microwave attenuation coefficient for electrodes. Additionally, MZM has ultra-high bandwidth also indicates well electro-optical velocity match. The microwave signal propagations only one way along transmission line from launch pad to 50 Ω termination, however, the optical signal travels forward and backward along the waveguide due to the reflection of loop mirrors in MIM, which leads to electro-optical velocity mismatch.

To obtain the performance of our MIM in high-speed digital data transmission, we performed on-off keying (OOK) modulation measurements. Figure 5 shows non-return-to-zero (NRZ) open eye diagrams at 32 Gbit/s and 40 Gbit/s. The 32 Gbit/s and 40 Gbit/s electrical signals are generated by an arbitrary waveform generation (AWG, Micram) with 64 Gb/s and 80 Gb/s sampling rate, respectively, then amplified by a linear amplifier (SHF 807) with peak-to-peak voltage ($V_{pp}$) of 6.1 V. Eye diagrams are obtained from a sampling oscilloscope (Tektronix 8300) without any electrical compensations. When the modulator is biased at the quadrature point, we measured the dynamic ERs are 8 dB at 32 Gbit/s and 6.6 dB at 40 Gbit/s are shown in Fig. 5(a) and (b) respectively. Our device operates at data rates higher than the measured 3 dB EO bandwidth because of the high electrical signal quality.

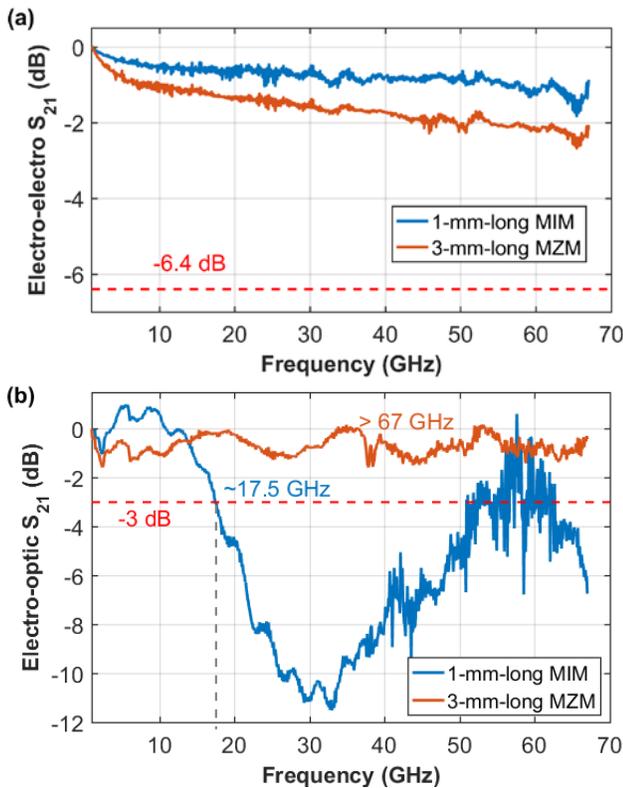

Fig. 4. The measured (a) electro-electro bandwidth ($S_{21}$ parameter) and (b) electro-optic bandwidth ($S_{21}$ parameter) of 1-mm-long MIM (~17.5 GHz) and 3-mm-long MZM (>67 GHz), dashed red lines indicate the 6.4 dB threshold of EE in (a) and 3dB threshold of EO in (b). $S_{21}$, transmission coefficient of the scattering matrix.

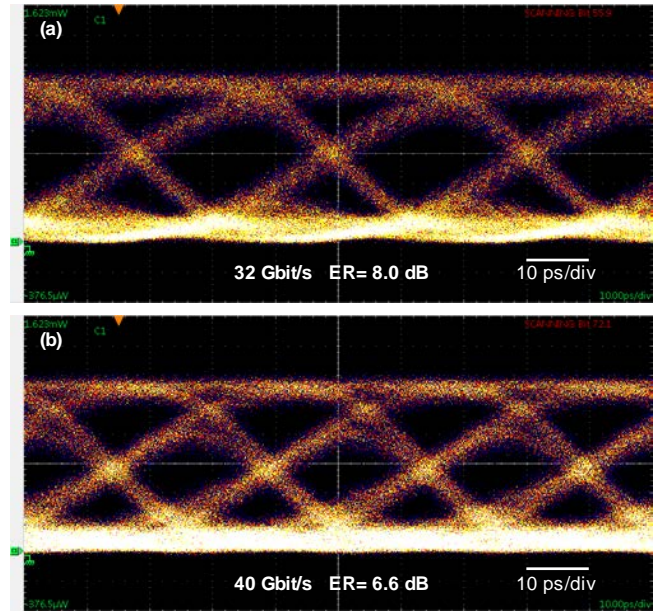

Fig. 5. The measured optical open eye diagrams at (a) 32 Gbit/s and (b) 40 Gbit/s under bias voltage at below the quadrature point. The dynamic extinction ratios are 8.0 dB and 6.6 dB, respectively.

In summary, we demonstrate a heterogeneous silicon and LN MIM in compact footprint (0.1 mm²) with doubled modulation efficiency compare to a MZM. The highest modulation efficiency $V_\pi \cdot L$ in LN platform of 1.2 V·cm is achieved. In the high-speed data transmission experiment, we have shown an open optical eye diagram data rates up to 40 Gbit/s with 6.6 dB extinction ratio, which is comparative to that of silicon modulators. The present device demonstrate an appealing insertion loss of around 3 dB. This hybrid platform can also avoids some disadvantages in conventional silicon modulators, such as absorption loss and nonlinearity. It should be noted that the present MIM showed a limited EO bandwidth, which is due to the velocities matching cannot be achieved for both propagation synchronously.

**Funding.** National Natural Science Foundation of China (NSFC) (11690031, 61575224, 61622510), Local Innovative and Research Teams Project of Guangdong Pearl River Talents Program (2017BT01X121)